\documentclass[%
 reprint,
 amsmath,amssymb,
 aps,superscriptaddress,
]{revtex4-2}

\usepackage{graphicx}
\usepackage{dcolumn}
\usepackage{bm}
\usepackage{braket}
\usepackage{xfrac}
\usepackage[separate-uncertainty = true, multi-part-units=single]{siunitx}
\usepackage{xcolor}
\usepackage{gensymb}
\usepackage{natbib}
\usepackage{chemformula}

\begin{document}

\preprint{APS/123-QED}

\title{Scalable quantum interference in integrated lithium niobate nanophotonics}

\author{Tristan Kuttner}
\email{tkuttner@ethz.ch}
\affiliation{Optical Nanomaterial Group, Institute for Quantum Electronics, Department of Physics, ETH Zurich, CH-8093 Zurich, Switzerland}

\author{Alessandra Sabatti}
\affiliation{Optical Nanomaterial Group, Institute for Quantum Electronics, Department of Physics, ETH Zurich, CH-8093 Zurich, Switzerland}

\author{Jost Kellner}
\affiliation{Optical Nanomaterial Group, Institute for Quantum Electronics, Department of Physics, ETH Zurich, CH-8093 Zurich, Switzerland}

\author{Rachel Grange}
\affiliation{Optical Nanomaterial Group, Institute for Quantum Electronics, Department of Physics, ETH Zurich, CH-8093 Zurich, Switzerland}

\author{Robert J. Chapman}
\email{rchapman@ethz.ch}
\affiliation{Optical Nanomaterial Group, Institute for Quantum Electronics, Department of Physics, ETH Zurich, CH-8093 Zurich, Switzerland}


\begin{abstract}

Photonics has emerged as one of the leading platforms for the implementation of real-world-applicable quantum technologies, enabling secure communication, enhanced sensing capabilities, as well as resolving previously intractable computational challenges. 
However, to harness the full potential of the photonics platform, several engineering feats need to be accomplished, among those is the quest for a scalable source of pure single photons.
While single photon sources can be implemented in a variety of different ways, integrated lithium niobate stands out as a prime contender for a monolithic quantum photonics platform, given its second-order nonlinearity and proven classical scalability. 
Despite the extensive effort put into developing the platform, integrating suitable photon pair sources remains a hurdle limiting the scalability of quantum photonic systems in lithium niobate.
We engineer three-wave-mixing in a nanophotonic lithium niobate device, integrating multiple near-perfect spectrally separable heralded single photon sources.
By mixing photons generated via the developed sources, we show bosonic interference between indistinguishable photons, a crucial interaction for many photonic quantum computing protocols.
This demonstration of the first proof-of-principle multi-source interference in integrated lithium niobate contributes to developing a truly scalable quantum photonics platform.

\end{abstract}

\maketitle

\subsection{Introduction}

Photonics stands out as one of the leading hardware implementations for quantum systems, offering an already existing advantage over their classical counterpart in various applications ranging from communications, to sensing and computing tasks \cite{obrien_photonic_2009, wang_integrated_2020}. 
Quantum communication leads as the most mature quantum technology, allowing for information-theoretic secure communication, with its main challenges being of economic and infrastructure related nature \cite{azuma_quantum_2023}. 
With a wide variety of applications, quantum sensors not only enhance precision measurement tools in our every day lives, but also enable the observation of new physical phenomena \cite{he_quantum_2023, jia_squeezing_2024}.
Achieving computational advantage using quantum technologies has the potential to transform our society in a vast range of areas. 
However, current state of the art quantum photonics experiments either lack the scale for useful applications \cite{yu_von-neumann-like_2024} or universality \cite{madsen_quantum_2022, zhong_quantum_2020}.
In order to bridge the gap to a practical quantum advantage there are numerous challenges along the way, from the discovery of less demanding, more robust, quantum protocols, all the way to the technological difficulties of improving and scaling up hardware implementations \cite{acharya_quantum_2025, moody_2022_2022, baboux_nonlinear_2023}.

\begin{figure*}[htbp]
    \centering
    \includegraphics[width=\textwidth]{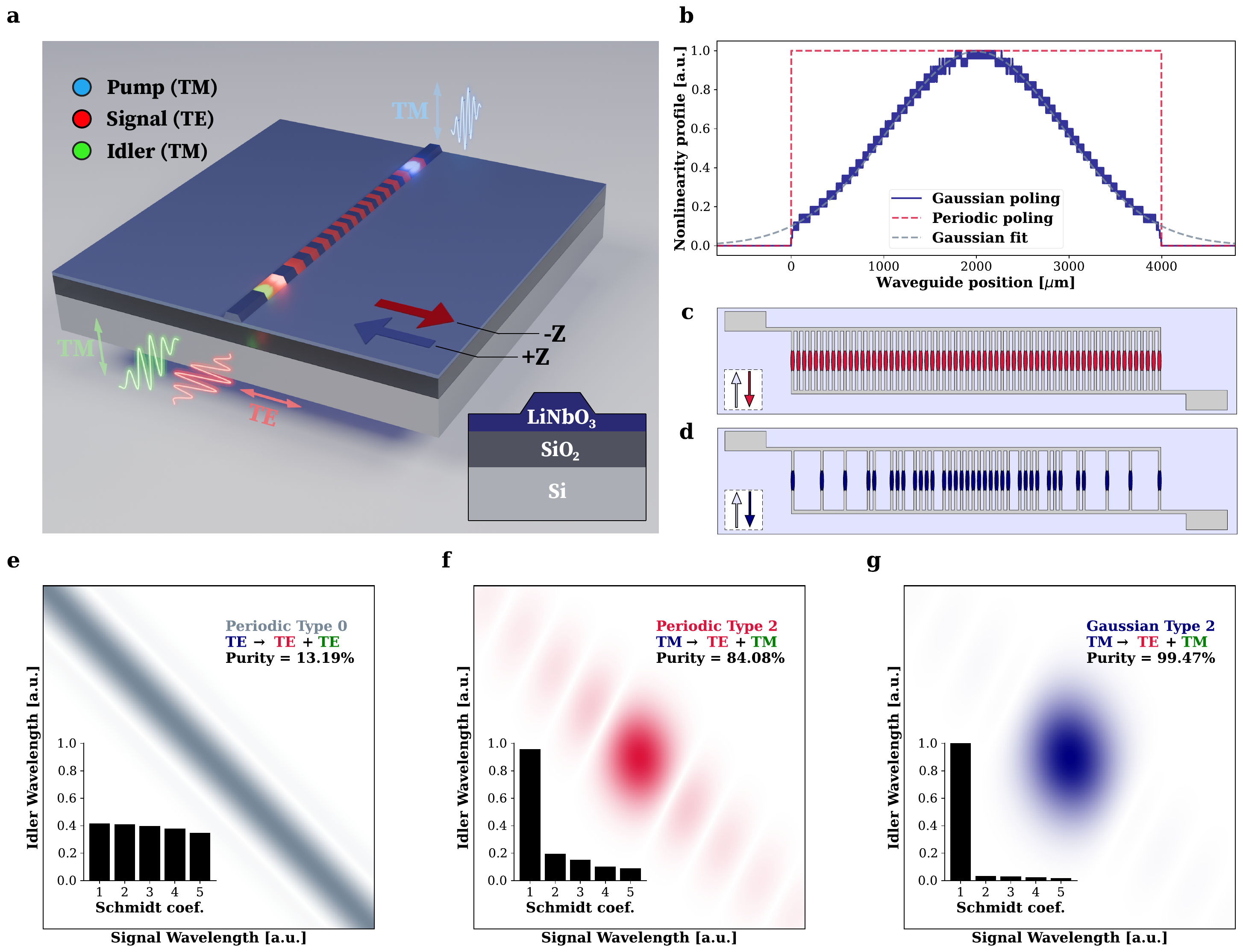}
    \caption{\textbf{a}, Schematic of an integrated Gaussian poled lithium niobate waveguide, downconverting a vertically TM (blue) polarized pump photon into a horizontal TE (red) signal and vertical TM (green) idler photon. The red and blue arrows indicate the orientation of the poled crystal. The inset shows the typical material stack for lithium niobate on insulator devices. \textbf{b}, Constant absolute nonlinearity profile for periodic poling (red), poling density approximating a Gaussian (blue), Gaussian fit of the approximation (gray), along the waveguide position. \textbf{c-d}, Schematic representation of poling electrodes (gray) for (\textbf{c}) periodic poling and (\textbf{d}) Gaussian poling using the deleted domain approach. \textbf{e}-\textbf{g}, Simulated joint spectral amplitude for: \textbf{e} periodically poled Type 0 SPDC, \textbf{f} periodically poled Type 2 SPDC, \textbf{g} Gaussian poled Type 2 SPDC.}
    \label{fig:concepts}
\end{figure*}

Due to the advancements in nanofabrication technologies over the past two decades \cite{rabiei_optical_2004}, thin-film lithium niobate on insulator (LNOI) integrated photonics, enabling a vast range of indispensable features, emerged as one of the premier contenders for a monolithical platform facilitating the implementation of large scale photonic quantum systems \cite{zhu_integrated_2021, pelucchi_potential_2022}. 
Among its favorable properties are a large index contrast and low material loss, allowing for the integration of dense, high fidelity photonic circuits \cite{shams-ansari_reduced_2022}. 
However, what sets LNOI apart from other advanced platforms like \ch{Si} or \ch{SiN} photonics is its large $\chi^{(2)}$ nonlinearity. 
This nonlinerity allows for three-wave mixing in LNOI, enabling two extremely attractive processes: the electro-optic Pockels effect and spontaneous parametric down-conversion (SPDC). 
Given the electro-optic phase tunability, bulk lithium niobate has already revolutionized the landscape of classical telecommunication, vastly accelerating data transmission over long distances.
Owing to the stronger mode confinement of integrated waveguides, even higher modulation speeds, greater than \SI{100}{GHz}, are achievable in LNOI \cite{zhang_integrated_2021}.
The capability to modulate the speed of light traveling along low-loss waveguides, and therefore reconfigure interferometric networks as well as arbitrary linear optical circuits, at exceedingly high frequencies, allows for the implementation of ultra high clock rate classical and quantum computation \cite{aghaee_rad_scaling_2025, lin_120_2024, wang_multidimensional_2018}.
High brightness SPDC sources in LNOI can be engineered by inverting the non-centrosymmetric crystal orientation of LN, by applying high voltage pulses in a process called poling, thus flipping the sign of the effective nonlinearity \cite{reitzig_seeing_2021, yamada_firstorder_1993, xin_spectrally_2022, sabatti_extremely_2024, fejer_quasi-phase-matched_1992}.  
When pumping these poled waveguides at a certain wavelength, a higher energy photon can be down converted into two lower energy photons, as schematically shown in Fig. \ref{fig:concepts}(a).
The probabilistically generated SPDC photon pairs can be used to generate the necessary resource states for a variety of quantum photonic applications \cite{knill_scheme_2001, bartolucci_fusion-based_2023, zhong_12-photon_2018, bentivegna_experimental_2015}.

Many of the promising quantum computing schemes rely on the interference of multiple heralded single photons from different sources, where one of the photons of each pair generated in the SPDC process is detected, to confirm the existence of the other one \cite{cao_photonic_2024, bentivegna_experimental_2015, bartolucci_fusion-based_2023, knill_scheme_2001}. 
This heralding process collapses the quantum state of the undetected photon into a specific projection, depending on the correlations between the two photons \cite{graffitti_design_2018}. 
Given the lack of a suitable Kerr-nonlinearity in photonics, multi-photon interference relies on their bosonic nature and the resulting bunching of indistinguishable photons \cite{hong_measurement_1987}. 
Consequently, it is necessary to generate separable SPDC photon pairs, in order to ensure indistinguishable heralded photons and thus, high fidelity quantum interference between them \cite{branczyk_hong-ou-mandel_2017}. 
In most LNOI SPDC implementations the generated photons are spectrally highly correlated, which renders them useless as a resource for a scaled quantum system relying on multiphoton interference \cite{zhao_high_2020, chapman_-chip_2025}.

In this work, we demonstrate the integration of multiple indistinguishable and spectrally uncorrelated heralded photon pair sources on a LNOI photonic integrated circuit and perform on-chip quantum interference.
We carefully engineer the waveguide dispersions, nonlinear profile via aperiodic Gaussian poling, and pump pulses of two independent sources, reaching estimated spectral purities of \SI{95.2}{\percent} and \SI{93.8}{\percent}. 
By interfering both the heralded photons at \SI{1505}{\nm} wavelength, we achieve \SI{70.72(5.02)}{\percent} quantum interference visibility, limited mainly by the on-chip beamsplitter reflectivity.
Making use of these sources, we demonstrate the first fully integrated proof-of-principle multi-source interference experiment in the LNOI platform, pushing the development of it as a truly scalable integrated quantum photonics platform.

\subsection{Results}
\subsubsection*{Pure integrated SPDC sources}

\begin{figure*}[htbp]
    \centering
    \includegraphics[width=\textwidth]{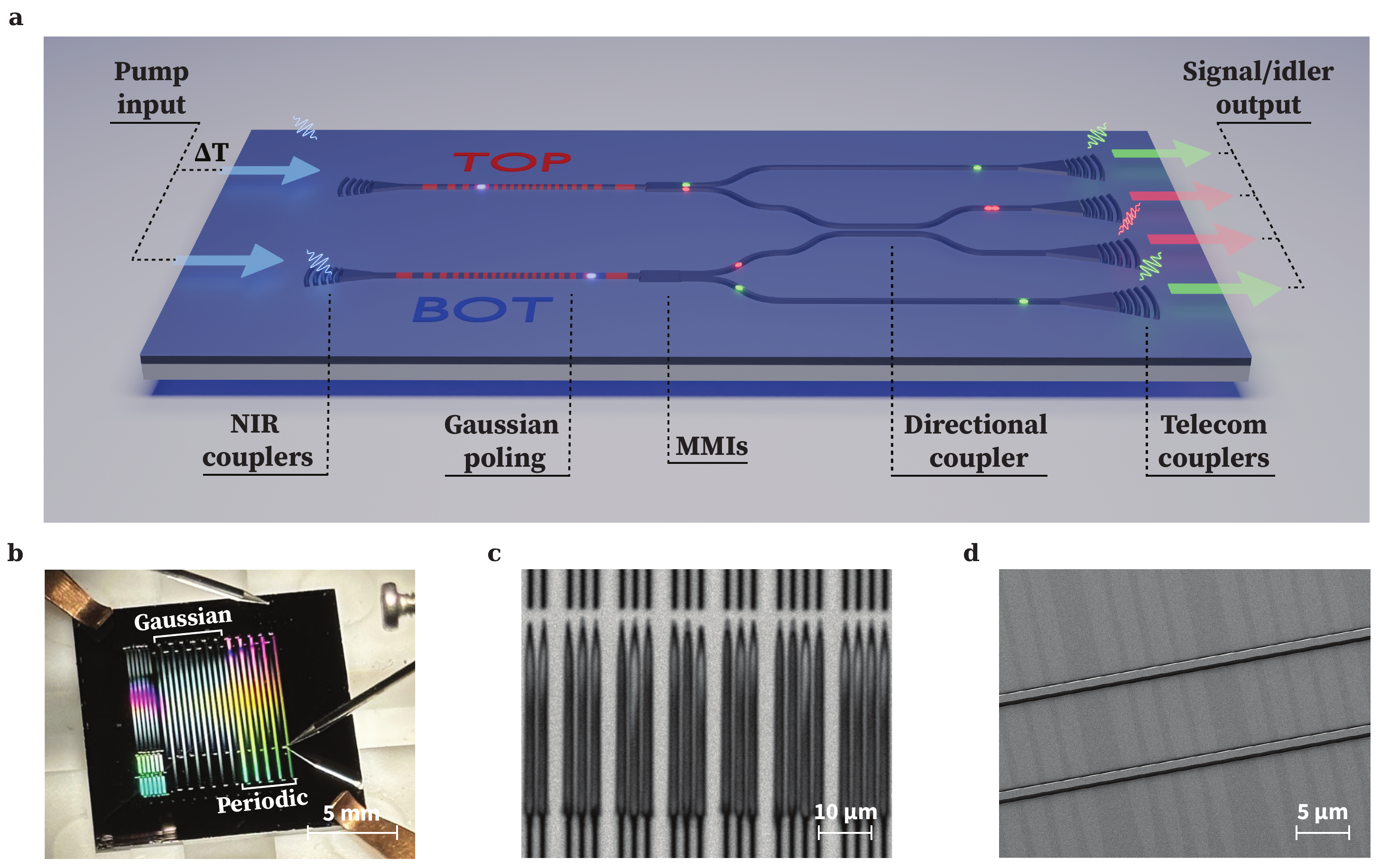}
    \caption{\textbf{a}, Schematic of the fabricated two-source interference chip. \textbf{b}, Poling process of the chip. \textbf{c}, Multiphoton microscopy image of the poled domains. \textbf{d}, Scanning electron microscopy image of two poled waveguides. }
    \label{fig:implementation}
\end{figure*}

The spectral purity of heralded biphoton states generated via SPDC is determined by the joint spectrum $f(\omega_s, \omega_i)=\alpha_p(\omega_s + \omega_i)\phi(\omega_s, \omega_i)$ of the generated signal and idler photons.
It is determined as a convolution of both the fundamental conservation principles involved in the process: energy conservation, described by the pump-envelope function (PEF) $\alpha_p(\omega_s + \omega_i)$, and momentum conservation, described by the phase-matching function (PMF) $\phi(\omega_s, \omega_i)$.
The PEF is characterized by the pump spectral and temporal properties, all parameters which are experimentally tunable.
The PMF in poled $\chi^{(2)}$ crystals, is mainly determined by the dispersion relations, represented by $\Delta k = k_p(\omega_s + \omega_i) - k_s(\omega_s) - k_i(\omega_i)$, of the involved pump, signal, and idler tones, as well as the spatial profile of nonlinear interaction strength $g(x)$ along the crystal length $L$, which is influenced by the poling period and distribution of poled regions.
    \begin{equation}
        \phi(\omega_s, \omega_i) = \int_0^L g(x)e^{i\Delta k(\omega_s, \omega_i)x} dx
        \label{eq:PMF}
    \end{equation}
By using the process of Schmidt decomposition, the joint spectrum of the generated photon pair state can be described as a sum of orthogonal modes:
    \begin{equation}
        f(\omega_s, \omega_i) =  \sum_n \nu_n \psi_n(\omega_s) \varphi_n(\omega_i)
        \label{eq:schmidt}
    \end{equation}
with $\psi_n(\omega_s)$ and $\varphi_n(\omega_i)$ being the Schmidt modes describing the signal and idler spectra and $\nu_n$ the Schmidt coefficients.
To generate pure heralded single photons, both the PEF as well as the PMF need to be engineered with respect to each other to allow for a separable joint spectrum $f(\omega_s, \omega_i) = \psi_0(\omega_s) \varphi_0(\omega_i)$, with only a single non-zero Schmidt coefficient \cite{graffitti_design_2018}.

The dispersion relations in integrated waveguides are mainly given by the material properties and fabricated geometries. 
For Type 0 and Type 1 phase matching, the generated signal and idler photons are in the same polarization, thus featuring the same dispersion properties.
In the telecom wavelength range, a process like this always generates spectrally highly correlated photon pairs, as illustrated in Fig. \ref{fig:concepts}(e), for Type 0 SPDC.
By allowing one of the generated photons to be of different polarization, they consequently also have different dispersion, thus enabling more freedom in tuning the dispersion relations between pump, signal, and idler. 
By carefully engineering the waveguide dimensions, we can tune the PMF to enable significantly higher spectral purity for photon pairs generated with Type 2 SPDC, as seen from the JSA in Fig. \ref{fig:concepts}(f).

The spatial nonlinearity profile $g(x)$ in equation \ref{eq:PMF} introduces an additional dependence of the joint spectrum of the generated photons on the structure of the poled crystal.
In the case of periodic poling, the nonlinear interaction strength stems from a top-hat like density of poled domains along the propagation direction of the light, as shown in Fig. \ref{fig:concepts}(b) with a schematic of the corresponding poled domains in Fig. \ref{fig:concepts}(c), which leads to a frequency response proportional to $\rm{sinc}(\Delta k \frac{L}{2})$.
Consequently, a sinc-like PMF introduces spectral correlations coming from its sidelobes, which renders it useless for the application as a scalable source of spectrally pure quantum states. 
To resolve this issue and remove correlations rooted in the spatial profile of the nonlinearity, we aim for a Gaussian-like frequency response, which can be achieved by engineering the spatial profile to also be Gaussian.
For this we approximate a Gaussian interaction strength, by modulating the density of poled domains with a Gaussian function \cite{tambasco_domain_2016}. 
We can see the Gaussian approximation of poled periods and a fit in Fig. \ref{fig:concepts}(b) as well as a schematic of poling electrodes used for this process in Fig. \ref{fig:concepts}(d).
By engineering the poling profile in this way, it is possible to strongly suppress the sidelobes in the JSA, as evident when comparing Figures \ref{fig:concepts}(e) and (f).
Combining this approach with a Type 2 process unlocks near-perfect spectrally separable SPDC in integrated LNOI sources.
Figure \ref{fig:concepts}(g) shows the simulated JSA for a dispersion engineered Gaussian poled Type 2 process, featuring near unity spectral purity.

Figure \ref{fig:implementation}(a) shows the schematic design and working principle of our implemented device: two separate high purity SPDC sources are independently pumped by a mode locked laser, then the generated photons are probabilistically split at a multimode interferometer (MMI). The TM polarized idler photons are coupled off the chip using polarization selective TM grating couplers optimized by inverse design. The TE signal photons are each sent to one of the inputs of a directional coupler, which acts as a beam splitter, and then coupled out using TE grating couplers.
The respective performances of passive devices are showcased in the Supplementary Material \ref{SUPP_S1}.

Our device is fabricated on a \SI{600}{\nano\meter} thick x-cut LNOI film, motivated by the dispersion engineering conditions leading to high spectral separability of Type 2 SPDC sources, as shown in Fig. \ref{fig:concepts}(g) and the Supplementary Material \ref{SUPP_S2} \cite{xin_spectrally_2022}. 
The Gaussian domain inversion of the sources is implemented by applying high voltage pulses to lithographically patterned electrodes on the chip.
Figure \ref{fig:implementation}(b) shows a photograph of the chip during poling and Fig. \ref{fig:implementation}(c) a multiphoton microscopy image of the inverted domains, with some missing according to the deleted domain approach \cite{tambasco_domain_2016}.
Following this process the waveguides structures are transfered onto the crystal via electron beam lithography and a physical ion etch, with the exact process flow layed out in the Methods section.
A scanning electron microscopy image of the final fabricated poled waveguides is shown in Fig. \ref{fig:implementation}(d).

\begin{figure*}[htbp]
    \centering
    \includegraphics[width=\textwidth]{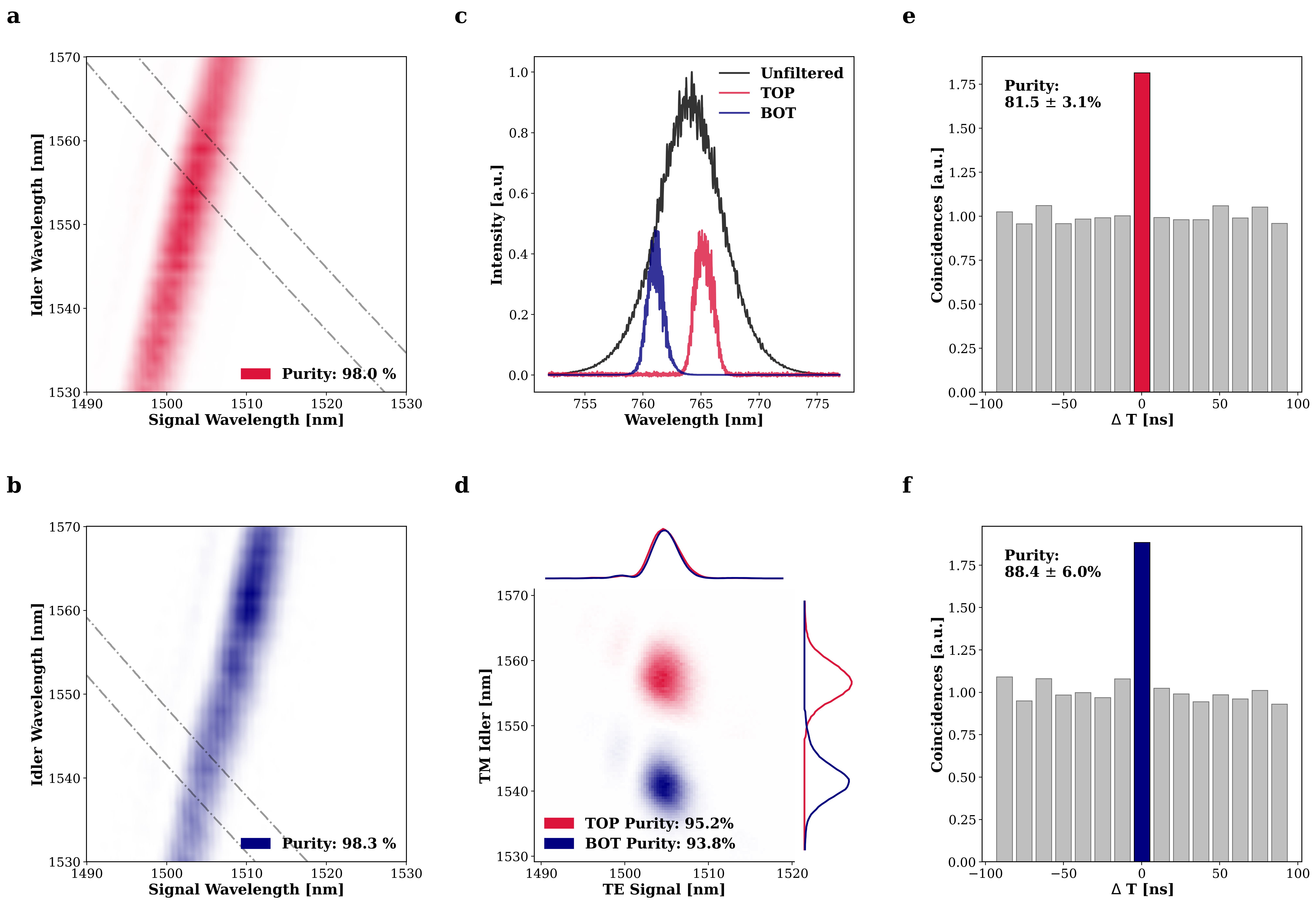}
    \caption{\textbf{a}-\textbf{b}, Map of the $|\mathrm{PMF}|^2$ of the TOP (BOT) source, with the gray dashed lines indicating the FWHM of a simulated PEF to estimate the source purity. \textbf{c}, Initial unfiltered \si{\femto\second} pulse (black) and filtered spectra for TOP (red) and BOT (blue) sources. \textbf{d}, Joint spectral intensity of TOP (red) and BOT (blue) source, with the projection onto the signal (up) and idler (right) axis. \textbf{e}-\textbf{f}, Autocorrelation measurement of the TOP (BOT) source.}
    \label{fig:results}
\end{figure*}

As an initial characterization we probe the PMF of both integrated sources by pumping the reverse process of SPDC, sum-frequency generation (SFG), where two tones of similar frequency are mixed and up-converted to a higher frequency. 
We couple two tunable continuous wave laser sources onto the chip, one via the TE signal grating and the other via rotating the polarization in fiber and the TM idler grating.
These two tones are combined at the MMI and pump SFG in the poled waveguides.
The generated near-infrared (NIR) signal is coupled off-chip, demultiplexed from any telecom contributions, and detected by a \ch{InGaAs} photodetector. 
A schematic of this measurement setup, and all the following ones, can be found in the Supplementary Material \ref{SUPP_S3}.
By sweeping signal and idler wavelengths we can measure the $\mathrm{|PMF|}^2$, as shown in Fig. \ref{fig:results}(a (b)) for the TOP (BOT) source, both of which feature strong sidelobe suppression. 
Given a slight shift in the phase-matching of the two sources, we need to pump their SPDC process with shifted spectra to ensure maximal spectral overlap of the signal photons. Assuming Gaussian pumps with optimized bandwidths for purity, and center wavelengths for spectral overlap, we can estimate the heralded state purity of the TOP (BOT) source from the $\sqrt{|\mathrm{PMF}|^2}$ to be \SI{98.0}{\percent} (\SI{98.3}{\percent}). 
For more details on the data acquisition and processing see the respective Methods section.

We use a \si{\femto\second} pulsed laser with a bandwidth of \SI{6.6}{\nano\meter} at a center wavelength of \SI{764}{\nano\meter} to pump both integrated sources. Using this bandwidth for the SPDC process would lead to strong spectral correlations, thus we carve out two different narrow bandwidth spectra of \SI{1.84}{\nano\meter} (\SI{1.65}{\nano\meter}) with center wavelength of \SI{765.2}{\nano\meter} (\SI{761.1}{\nano\meter}) for the TOP (BOT) source, as shown in Fig. \ref{fig:results}(c), to ensure optimal bandwidth matching between PEF and PMF, as well as spectral overlap of the signal photons.

To directly probe the joint spectral intensity (JSI) $|f(\omega_s, \omega_i)|^2$ we pump each of the sources with their respective carved pump spectra, and perform time-of-flight spectroscopy on the generated photon pairs.
For this measurement the outcoupled tones are cleaned from any remaining signal/idler leakage by filtering with a set of fiber coupled polarization beamsplitters (PBS) and cleaned from any remaining pump photons by passing a NIR longpass filter. The signals and idlers are sent through a set of circulators and a spool of high dispersion fiber, equalling the negative dispersion of \SI{30}{\kilo\meter} SMF28 fiber. The strongly dispersed signal and idler photons are detected using superconducting nanowire single photon detectors (SNSPD), gated by the trigger output of the pump laser.
The temporal correlations between arriving signal and idler photons are recorded with a time tagging device, and then mapped back to spectral correlations, given given the known dispersion of the fiber.
In Fig. \ref{fig:results}(d) we see the TOP (BOT) JSI in red (blue), and the projections of both JSIs onto the signal and idler axis, validating the choice of respective pump spectra for the different sources.
We estimate the heralded spectral purity from the $\mathrm{\sqrt{JSI}}$ of the TOP (BOT) source to be \SI{95.2}{\percent} (\SI{93.8}{\percent}) \cite{graffitti_design_2018}.

Lastly, we characterize the sources by measuring their unheralded autocorrelation $g^{(2)}(\Delta T)$, which lets us determine the purity as $P \approx \frac{g^{(2)}(0)}{g^{(2)}(\infty)} - 1$ \cite{christ_probing_2011}. 
In the same way as for the time-of-flight spectroscopy we couple off the idler photons and clean them from any remaining pump. Additionally, we use a PBS to filter by polarization as well as employing an additional telecom broadband bandpass filter to make sure all signal photons are removed.
The idler photons are then fed into a fiber beamsplitter, with a working point near \SI{1550}{\nano\meter} motivating the choice of using the idlers, and correlations between the two outputs of the splitter are recorded via SNSPDs and a time tagger. 
Figure \ref{fig:results}(e (f)) show the autocorrelation measurements of the TOP (BOT) source, and from that a estimated purity of \SI{81.5(3.1)}{\percent} (\SI{88.4(6.0)}{\percent}), where we attribute the reduced purities to phase correlations introduced by chirp on the pump pulse as well as non-optimal poling.

\subsubsection*{Heralded quantum interference}

\begin{figure*}[htbp]
    \centering
    \includegraphics[width=\textwidth]{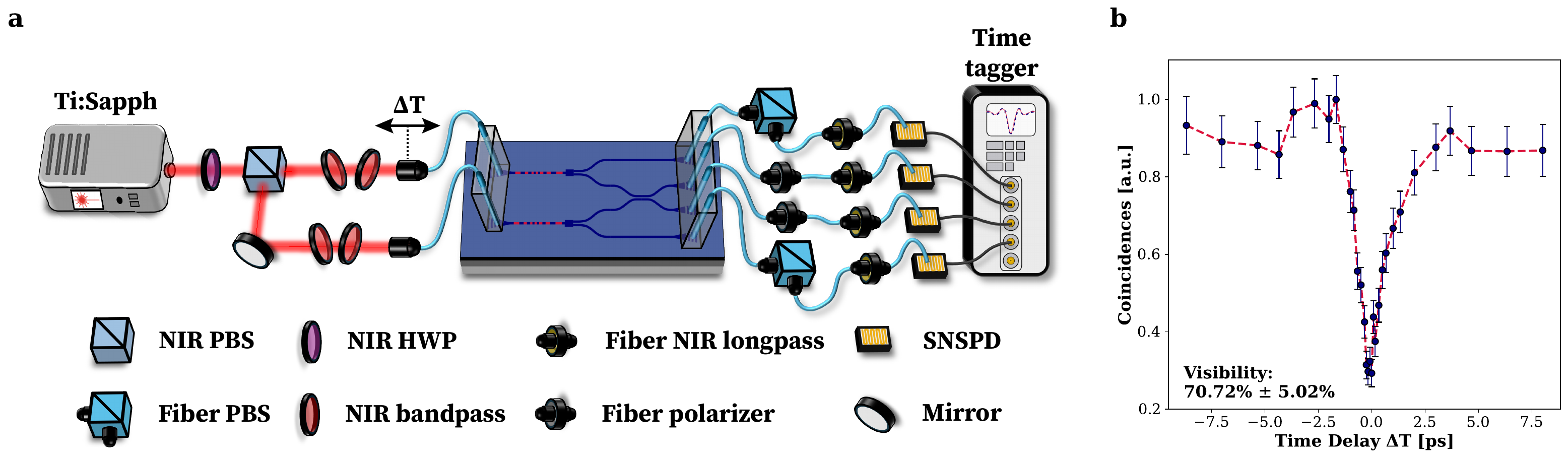}
    \caption{\textbf{a}, Schematic of the setup used to measure two-source quantum interference, with an artificial delay $\Delta T$ between the pump pulse for TOP and BOT source.
    \textbf{b}, Plot of the Hong-Ou-Mandel interference of the signal photons, exhibiting a reduction of the coincidence rate between its maximum and minimum of \SI{70.72(5.02)}{\percent} at $\Delta T = 0$. 
    } 
    \label{fig:3}
\end{figure*}

Due to the lack of a usable Kerr-nonlinearity for multi-qubit operations, many optical quantum computing schemes rely on the bosonic nature of photons, bunching at a beamsplitter, to implement a quasi-nonlinearity \cite{hong_measurement_1987}.
This bunching only happens if the photons are indistinguishable in every way, be it polarization, their temporal mode, or their spectrum. 
Photon pair sources based on SPDC lack deterministic emission, however, by detecting the presence of one of the photons it is possible to herald the existence of the other one.
In order to repeatedly generate pairs of indistinguishable heralded photons, no matter what spectral properties the detected photons have, the remaining ones should not be affected by this heralding process, thus they need to be spectrally separable.

To pump separate SPDC sources, we split the output of a Ti:Sapphire laser into two parts which are filtered to different spectra, as seen in Fig. \ref{fig:results}(c), and respectively coupled onto the TOP and BOT sources of the device.
When coupling the pulses, we use fiber collimators on movable stages to change the arrival time of the pulses at the SPDC sources, which in turn introduces a time offset $\Delta T$ onto the temporal modes of the generated photon pairs.
This temporal offset will be our knob to tune the distinguishability between the photon pairs.
Given the nature of integrated waveguides, the polarization selectivity of the grating couplers used to couple both signal and idler photons, as well as additional fiber polarization filters placed at the output of the chip, we can assume only those probabilistically split events where both TM idlers are in the heralding arms and TE signals are in the interfering arms, are accounted for.
We record fourfold coincidences between the outputs of the device while sweeping $\Delta T$, to observe quantum interference between the heralded signal photons.
Quantitatively, we expect a Hong-Ou-Mandel dip, with visibility V:
    \begin{align}
        V = & \lim_ {\tau \to 0} \frac{2R(1-R)}{R^2 + (1-R)^2} \sum_{n, n'} \nu^2_n \upsilon^2_{n'} \nonumber \\ 
        &\int d\omega_1 \psi_n^*(\omega_1)\varphi_{n'}(\omega_1) e^{-i \omega_1 \tau} \nonumber \\ 
        &\int d\omega_2 \varphi_{n'}^*(\omega_2)\psi_{n}(\omega_2) e^{i \omega_2 \tau}
        \label{eq:visibility}
    \end{align}
with R the reflectivity of the beamsplitter, $\nu$ and $\upsilon$ the Schmidt coefficients, and $\psi$ and $\varphi$ the respective Schmidt modes \cite{hong_measurement_1987, branczyk_hong-ou-mandel_2017, babel_demonstration_2023}.

Using an experimental setup as shown in Fig. \ref{fig:3}(a) we observe a visibility of \SI{70.72(5.02)}{\percent}, demonstrating multisource interference between two high purity integrated heralded single photon sources, with the full trace of measurements shown in Fig. \ref{fig:3}(b).
Evidently from equation \ref{eq:visibility}, not only the purity but also the spectral overlap of the photons, as well as the reflectivity of the used beamsplitter have a significant impact on the interference visibility. 
The integrated beamsplitter in this experiment has a reflectivity of $R=$ \SI{62.5}{\percent}, which reduces the visibility to $V=$ \SI{88}{\percent} under otherwise perfect conditions with spectrally completely separable photons.
A more comprehensive demonstration of the interference visibility with respect to the reflectivity can be seen in the Supplementary Material \ref{SUPP_S4}.
Additional factors reducing visibility are non-perfect spectral overlap of the signals, higher order SPDC events due to a high downconversion probability per pulse of around \SI{2.91}{\percent} (\SI{2.82}{\percent}), chirp of the pump pulses inducing phase correlations, as well as non-perfect pump bandwidth matching to the PMF and between the sources.
The shown interference visibility can thus be further increased by adapting the device layout to feature lower loss coupling elements, a tunable interferometer, as well as only a single free space coupled input, which was knowingly opted out of, to demonstrate the interference dip, in contrast to implementing the interference fixed at its maximum visibility with no temporal indistinguishability, which would be the approach when using the interference for computation schemes.

\subsection{Discussion}

In conclusion, we integrated multiple high-purity photon pair sources in the LNOI platform by engineering the waveguide dispersion as well as the nonlinearity profile of the waveguides, achieving among the highest spectral separability reported for LNOI based sources.
By making use of these photon pair sources we demonstrate the first proof-of-principle, fully integrated, quantum interference between independent heralded sources in the LNOI platform.
To further increase the source quality, refined fabrication processes can be employed to compensate for film thickness variations along the die \cite{chen_adapted_2023}, which would also allow for a longer phase-matching region leading to a reduced bandwidth of the generated photons.
Additionally, improving the poling to closer approximation a Gaussian can further suppress sidelobes and in turn enable higher heralded state purity \cite{graffitti_pure_2017}.

A scalable quantum photonics platform requires an extensive number of conditions, among those is the need for ultra low-loss coupling and detection efficiencies, fast and high fidelity reconfigurability and signal routing, cryogenic compatibility as well as spectrally pure photon sources.
The implemented highly spectrally separable photon pair sources make a contribution to manifest LNOI as a truly scalable quantum photonics platform, with potential use cases ranging from quantum computing, such as boson sampling, over communication, up to quantum state generation \cite{walther_experimental_2005, zhong_quantum_2020}.

\subsection{Methods}

\subsubsection*{Sum-frequency-generation}

The PMF of both sources of the device is measured via sum-frequency-generation (SFG), therefore the Gaussian poled waveguides are pumped with both TE and TM polarized light at telecom wavelengths. 
We couple onto the device using polarization sensitive grating couplers, which are illuminated by PM (PM15-U25D) fiber-arrays with a spacing of \SI{127}{\micro\meter} and a horizontal orientation of their slow axis. 
The TE signal is directly coupled onto the chip from a tuneable telecom laser (Keysight N7776C). The TM idler is coming from a different tunable telecom laser (Toptica CTL 1500), where the polarization inside the PM fiber is rotated via a fiber-coupled PBS (OzOptics). 
The generated SFG tone is coupled of the chip using a single SMF28 fiber, demultiplexed from any telecom signal remaining with a 780/1550 WDM (OzOptics), and detected with an \ch{InGaAs} photodetector (Keysight N7744C) synchronized to the signal laser. 
The wavelength of the signal is swept, with a step size of \SI{10}{\pico\meter}, while keeping the wavelength of the idler constant, after each signal sweep the wavelength of the idler is changed by \SI{1}{\nano\meter} and the signal sweep is repeated to sample the PMF map.
The resulting output map is post processed using a linear savgol filter with window width of \SI{1}{\nano\meter}.
When estimating the heralded purity of a source, given the measured $\mathrm{|PMF|^2}$, we take the square root of the measured data, before performing singular value decomposition, to estimate the purity from the $\sqrt{\mathrm{JSI}}$, which was shown to be the more accurate estimation \cite{graffitti_design_2018}.

\subsubsection*{Time-of-flight spectroscopy}

Time-of-flight spectroscopy of the SPDC process is performed in order to measure the JSI of the generated photons.
A \SI{80}{\MHz} \si{\fs} pulsed Ti:Sapphire laser (Spectra Physics Mai Tai HP DS) at a center wavelength of \SI{764}{\nm} with bandwidth of \SI{6.6}{\nm} is spectrally filtered by two narrowband bandpass filters (ThorLabs FBH780-10).
Changing the angle of incidence (AOI) of these filters induces a shift in their transmission spectra, enabling us to filter with a reduced bandwidth by using different AOIs for the filters.
The filtered pump pulse is coupled into fiber using a fiber collimation package and then onto the chip via a 780PM fiber array, with the pump light's polarization in the fibers parallel to the propagation direction on the chip, in order to couple to a TM mode of the integrated waveguide.
The TE signal and TM idler photons are probabilistically separated on-chip with an MMI, and filtered by means of polarization selective grating couplers to ensure the signal and idler exit the chip from the correct ports into a telecom PM fiber array.
To make sure there is no mixing of signal and idler in any of the outcoupled fibers, the signal output is filtered with an additional fiber polarizer and the idler output is filtered and rotated to the slow axis, with a fiber PBS.
Removing any remainders of the pump light is done via fiber coupled \SI{830}{\nano\meter} longpass filters (Semrock BLP01-830R).
The signal and idler tones are heavily dispersed in dispersion compensating fiber equaling about \SI{30}{\km} of SMF28 fiber, resulting in a temporal wavelength spacing of about \SI{0.5}{\nano\second\per\nano\meter}.
These dispersed pulses are then detected by superconducting nanowire single photon detectors (SNSPD) (Single Quantum Eos CS).
The detection events are recorded with a time tagging device (Swabian Time Tagger Ultra) and gated onto the arrival of the output trigger of the pulsed laser.

\subsubsection*{Autocorrelation measurement}

The second-order correlation fuction $g^{(2)}$ of the idler photons is measured with a standard Hanbury Brown and Twiss setup.
A similar coupling and filtering setup as for the joint spectrum measurements is used, however instead of dispersing the tones, the idlers are filterd by wideband spectral filters (Edmund optics TechSpec Bandpass Filter 1550/50nm) to remove any leakage of signal, then fed into a fiber splitter whose outputs are connected to two SNSPDs.
The time tagging device records the correlations between these two detectors, with the electronic delay tuned to have detection events from the same pulses at the same time.

\subsubsection*{Heralded Hong-Ou-Mandel interference}

To meassure the Hong-Ou-Mandel interference we split the pump laser with a half-wave plate and a PBS.
In each of the paths of the pump two BPFs are used to filter the spectrum, this enables us to carve two separate, narrowband, pulses out of the initial pump.
These separate pulses are coupled into fiber collimation packages, where one of them is mounted on a motorized linear translation stage (ThorLabs MT1/M-Z8).
The 780HP fiber array, used to couple the different pulses onto the chip, has its two fibers mounted with orthogonal axis with respect to each other. Thus, we couple the pulses into the fast axis of one of the fibers and the slow axis of the other one to ensure proper TM coupling onto the chip.
The two pulses each pump a separate SPDC process in two different sources, where the linear stage can introduce a time delay $\Delta T$ between these processes.
Each of the photon pairs is probabilistically split using an MMI.
The TM idler photons are directly coupled off the chip, the TE signal photons are routed to the two input ports of a directional coupler.
The two carved out pump pulses are choosen to match the respective PMFs in a way which produces TE Signal photons with as high as possible spectral overlap, their arrival time $\Delta T$ is their only distinguishing feature.
When we sweep the time delay between the arrival of the photons to be $\Delta T = 0$, they are, indistinguishable in every basis, thus, given their bosonic nature, bunch at either of the outputs of the beamsplitter.
If we change the arrival time of the photons to be $\Delta T\gg t_{c}$, with $ t_{c}$ being the coherence length of the photons, they will not preferably bunch or anti-bunch at the beamsplitter.
The signal photons are also coupled of the chip and filtered to their correct polarization, additionally all outcoupling channels feature \SI{830}{\nano\meter} longpass filters, to remove any remaining pump light. 
Finally, all four channels are routed to a SNSPD system to detect the single photons, where the arrival times and in particular fourfold coincidence events are recorded via a time tagging device.
In the case of similar arrival time $\Delta T \to 0$ we observe a reduction of the four-fold detection rate, which indicates Hong-Ou-Mandel interference.
The rate of four-fold coincidences is normalized with respect to the drifting downconversion probability of each of the sources, due to the drift in coupling when moving the translation stage. The detailed calculation of the downconversion probabilities can be seen in the Supplementary Material \ref{SUPP_S5}.

\subsubsection*{Device Fabrication}

The presented sample is fabricated out of a commercial \SI{600}{\nano\meter} lithium niobate film on a \SI{2}{\um} buried oxide layer (NANOLN). 
First, we pattern chromium poling electrodes using electron-beam lithography and a metal lift-off process.
We apply high voltage pulses to these electrodes to create a localized domain inversion of the ferroelectric \ch{LiNbO3} crystal, which we can inspect using a commercial confocal multiphoton microscopy system (Leica SP8 MP).
After stripping the electrodes, an electron beam lithography step is performed to pattern the waveguide structures onto a negative hydrogen silsesquioxane (HSQ) based resist.
These patterns are transfered into the LN thin film via physical ICP-RIE \ch{Ar+} etching of \SI{400}{\nano\meter} depth.
To remove the redeposited byproducts of the etching and the remaining mask, a  \ch{KOH} and a \ch{HF} based cleaning step is performed.

\subsection*{Data availability}

The raw data presented in this study is available from the authors upon reasonable request.

\subsection*{Competing interests}

The authors declare no competing financial or non-financial interests.

\subsection*{Author Contributions}

T.K. designed and simulated the device. T.K. and A.S. fabricated the lithium niobate-on-insulator sample including poling, lithography, etching and metal deposition. T.K. performed the optical measurements, data analysis and wrote the original draft of the manuscript. T.K., J.K., and R.J.C developed the theoretical framework of the experiment. R.J.C and R.G. supervised the project. All authors contributed to revising the manuscript.

\subsection*{Acknowledgments}

We acknowledge support for characterization of our samples from the Scientific Center of Optical and Electron Microscopy ScopeM and from the cleanroom facilities BRNC and FIRST of ETH Zurich and IBM Ruschlikon.
We acknowledge Danut-Valentin Dinu for support with the simulation of nonlinear frequency conversion processes and the setup of the dispersive fiber for time-of-flight spectroscopy.
R.J.C. acknowledges support from the Swiss National Science Foundation under the Ambizione Fellowship Program (Project Number 208707).
R.G. acknowledges support from the European Space Agency (Project Number 4000137426) and the Swiss National Science Foundation under the Sinergia Program (Project Number 206008).

\clearpage
\onecolumngrid 

\begin{center}
\textbf{\large Supplementary Material}
\end{center}
\vspace{1em}

\section{Classical device characterization}
\label{SUPP_S1}

\subsection{Grating couplers}

\begin{figure*}[htbp]
    \centering
    \includegraphics[width=1\textwidth]{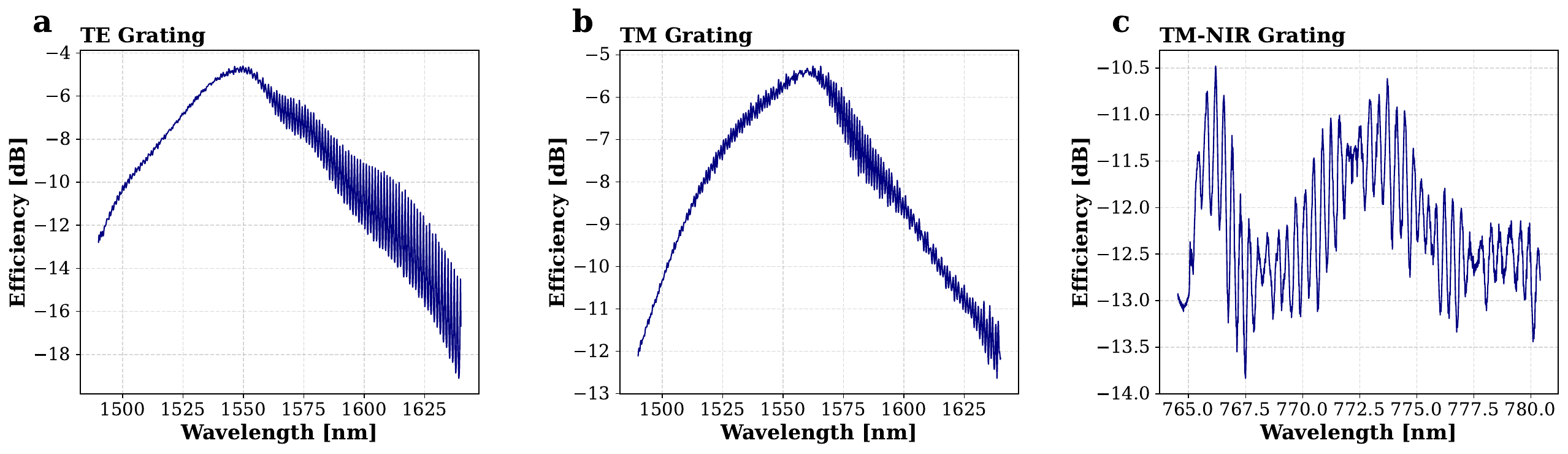}
    \caption{\textbf{a}, Coupling efficiency of the used telecom TE grating. \textbf{b}, Coupling efficiency of the used telecom TM gratings. \textbf{c}, Coupling efficiency of the used near-infrared TM gratings.}
    \label{fig:S3}
\end{figure*}

Coupling the TE signals, TM idlers and TM pump onto and off the chip requires different grating couplers.
We use a inverse design approach to optimize the coupling efficiency of those used couplers. Figure \ref{fig:S3} shows the respective efficiencies for the telecom TE and TM couplers as well as the near-infrared TM couplers.

\subsection{Directional couplers}

\begin{figure*}[htbp]
    \centering
    \includegraphics[width=0.67\textwidth]{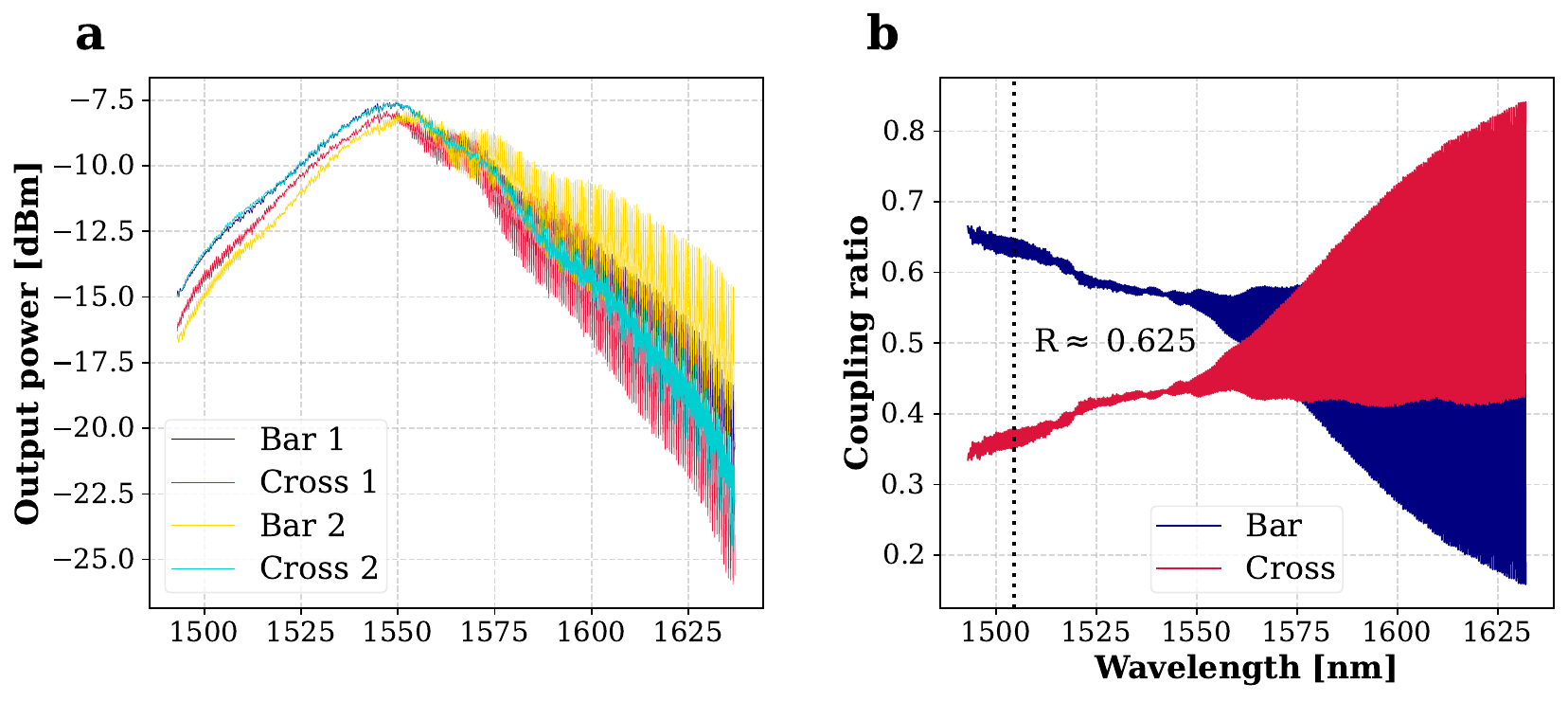}
    \caption{\textbf{a}, Bar and cross output of a test directional coupler for both inputs. \textbf{b}, Normalized bar and cross values of the device.}
    \label{fig:S4}
\end{figure*}

The splitting ratio of the integrated directional coupler mixing the signal photons from two different SPDC sources is wavelength dependent. Figure \ref{fig:S4}(a) shows the transmission when measuring a test coupler in transmission and reflection from both inputs. Figure \ref{fig:S4}(b) shows the normalized splitting ratio of the coupler, with the contribution of the gratings removed, it yields a reflectivity of $R=62.5$ at a wavelength of \SI{1505}{\nano\meter}, which is the averaged center wavelength of the signal photons used in the two source interference experiment. The measurement becomes less reliable at higher wavelengths due to stronger reflections at the grating couplers creating cavities.

\subsection{Multimode interferometers}

\begin{figure*}[htbp]
    \centering
    \includegraphics[width=0.33\textwidth]{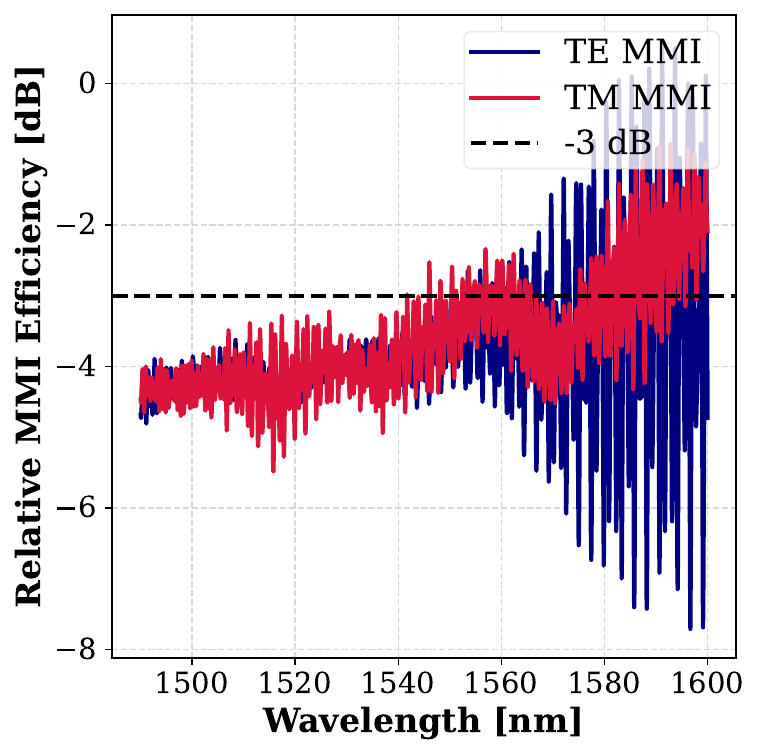}
    \caption{Transmission of a test MMI, normalized to the grating coupler efficiency, for TE and TM telecom signals.}
    \label{fig:S5}
\end{figure*}

The Multimode interferometers used in the device to probabilistically split the signal and idler photons each in a respective waveguide are characterized with a test device. The measured transmission for TE and TM light is normalized to the respective grating coupler efficiencies. The wavelength regions where the efficiency is greater than -3 dB can be attributed to different coupling efficiencies of the grating and MMI test devices.
\newpage

\section{Phase-Matching function simulation}
\label{SUPP_S2}
\begin{figure*}[htbp]
    \centering
    \includegraphics[width=0.67\textwidth]{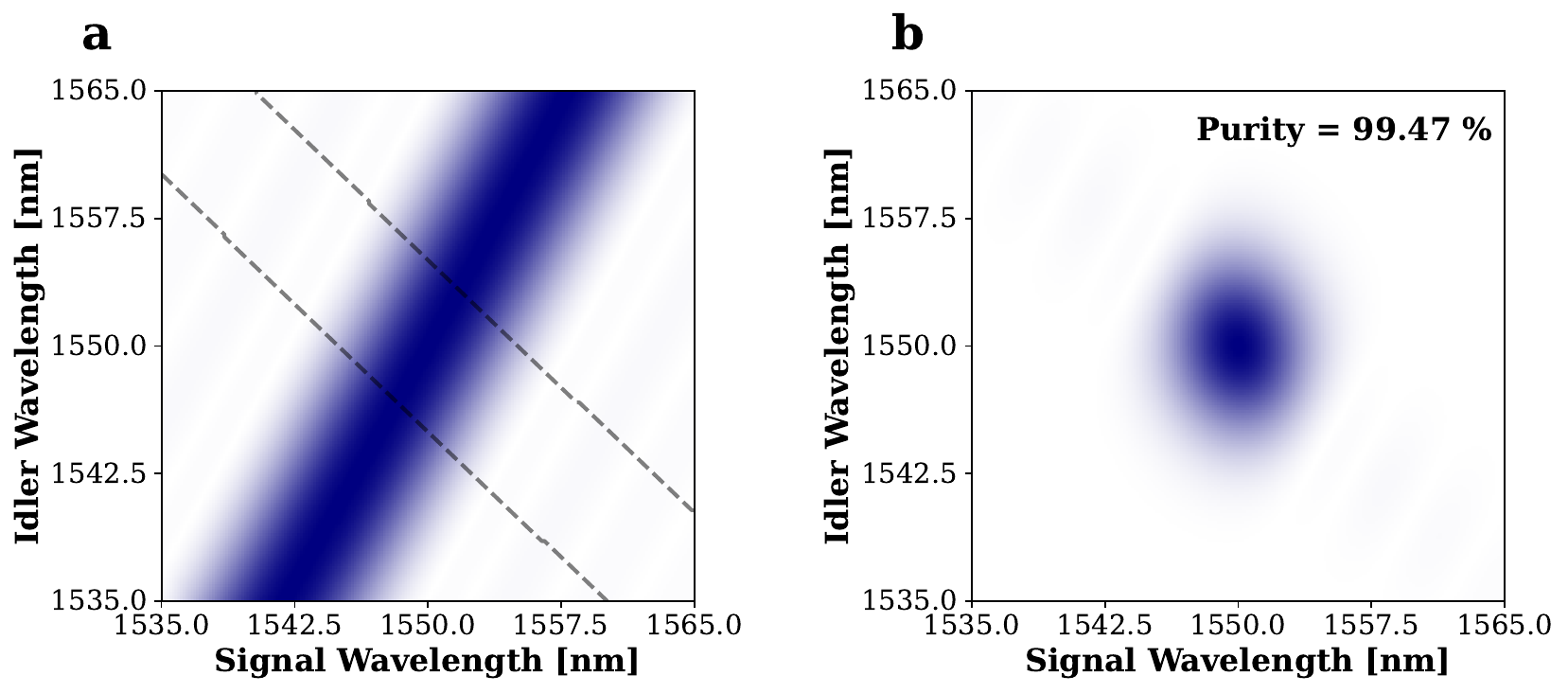}
    \caption{\textbf{a}, Simulation of the phase matching function of the design fabricated in this work. \textbf{b}, Estimated JSA with an assumed Gaussian PEF where the gray-dashed lines in (\textbf{a}) indicate it's FWHM.}
    \label{fig:S6}
\end{figure*}

We simulate the PMF of a Type 2 SPDC process in a Gaussian poled waveguide with our intended waveguide dimensions (see Fig. \ref{fig:S6}(a)). Assuming a Gaussian PEF with optimized bandwidth, we can estimate a JSA, which would lead to a heralded single photon purity of up to 99.47 \% (see Fig. \ref{fig:S6}(b)).

\newpage

\section{Experimental setups}
\label{SUPP_S3}
\begin{figure*}[htbp]
    \centering
    \includegraphics[width=0.67\textwidth]{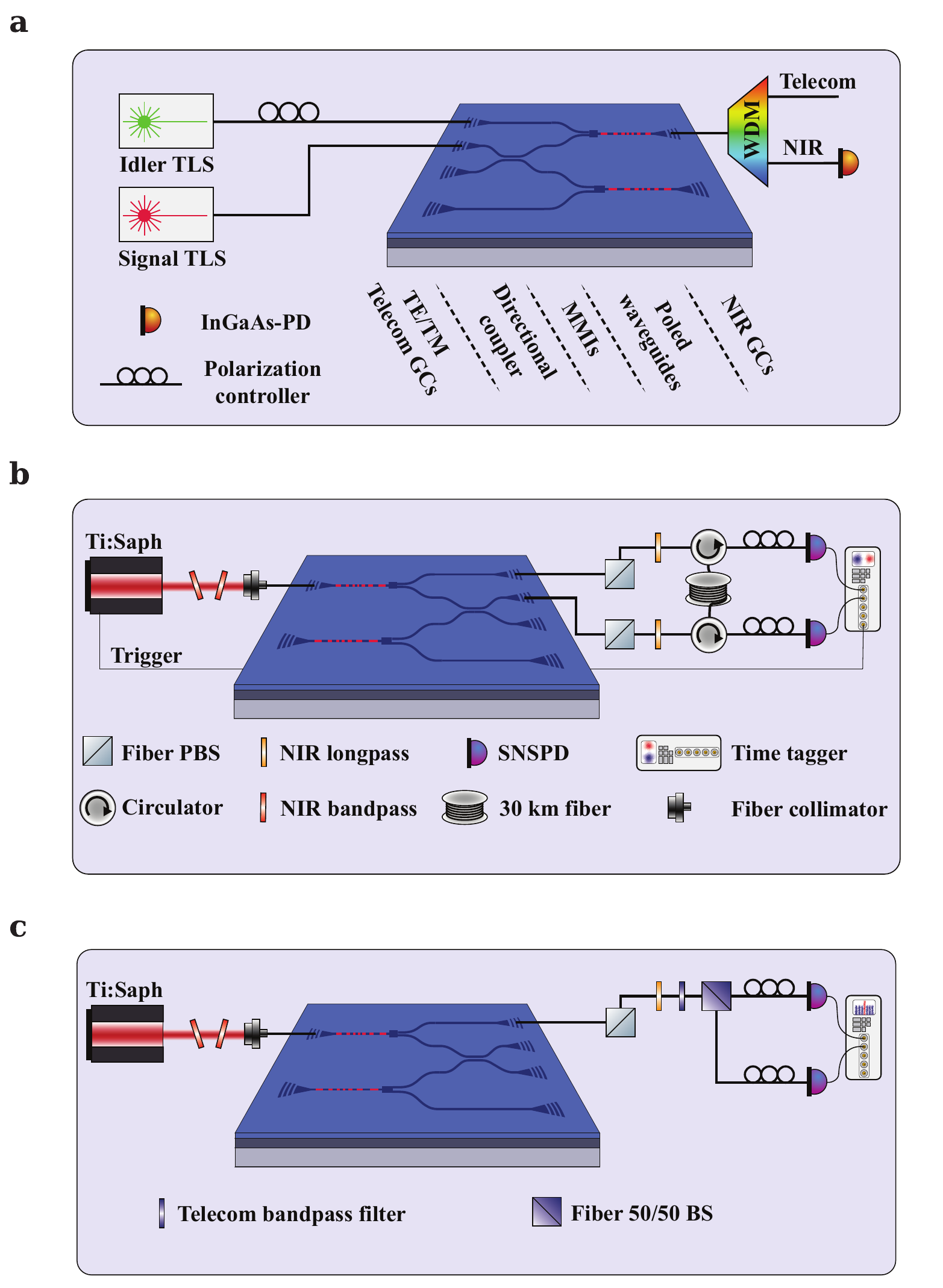}
    \caption{\textbf{a}, Schematic of the setup used to measure the SFG spectrum of the sources. \textbf{b}, Schematic of the setup used for the time-of-flight spectroscopy measurement. \textbf{c}, Schematic of the setup used to measure the autocorrelation of the generated SPDC photons.}
    \label{fig:S_Setups}
\end{figure*}

Each of the characterization measurements performed required a unique experimental setup. Figure \ref{fig:S_Setups}(a-c) show the setups for the SFG, time-of-flight, as well as autocorrelation measurements. Please refere to the Methods section for more details on the used components.

\newpage

\section{Impact of beamsplitter reflectivity on heralded HOM interference}
\label{SUPP_S4}
\begin{figure*}[htbp]
    \centering
    \includegraphics[width=0.33\textwidth]{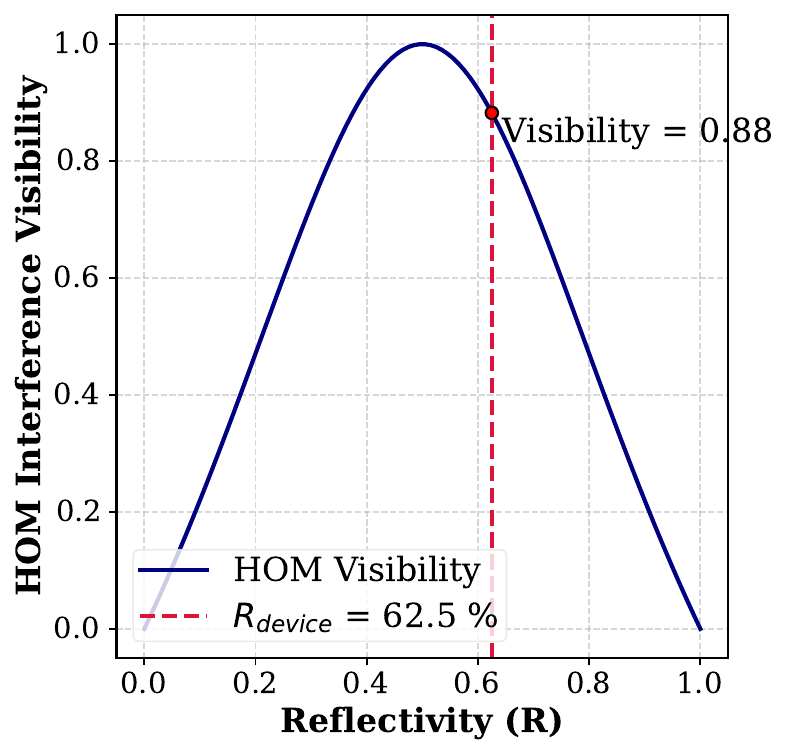}
    \caption{Impact of the reflectivity of a beamsplitter on the Hong-Ou-Mandel interference visibility under otherwise optimal conditions. The dashed vertical line represents the reflectivity of the integrated directional couplers used in this work. }
    \label{fig:S7}
\end{figure*}

The Hong-Ou-Mandel interference visibility between two indistinguishable photons is strongly influenced by the reflectivity of the used beamsplitter. We simulate the maximal visibiliy of the interference assuming optimal conditions with pure, indistinguishable single photons, but a varying reflectivity of the beamsplitter. The results can be seen in Fig. \ref{fig:S7}, where we also indicate the maximal visibility of 88\% for an experiment using the directional couplers we use, with a reflectivity of $R=0.625$.

\newpage

\section{Downconversion probability per pulse}
\label{SUPP_S5}

\begin{figure*}[htbp]
    \centering
    \includegraphics[width=1\textwidth]{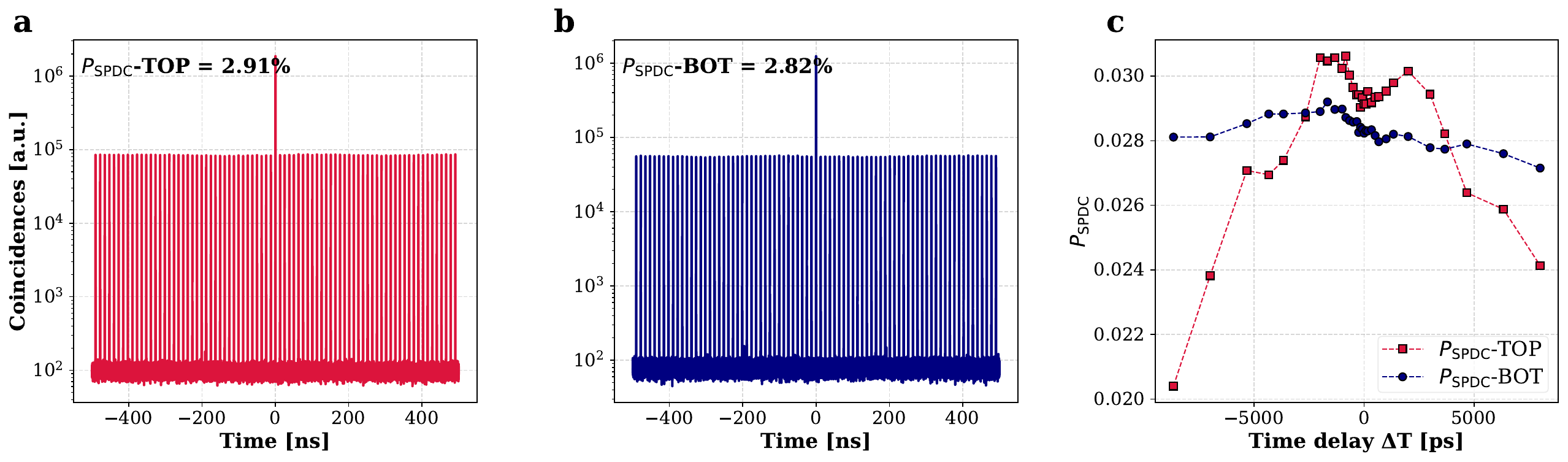}
    \caption{\textbf{a}, Bidirectional histogram of idler and BAR signal outputs of the TOP source, at a stage position equaling $\Delta T=0$, leading to a calculated downconversion probability of $P_{SPDC, TOP} = 2.91 \%$. \textbf{b}, Bidirectional histogram of idler and BAR signal outputs of the BOT source, at a stage position equaling $\Delta T=0$, leading to a calculated downconversion probability of $P_{SPDC, BOT} = 2.82 \%$. \textbf{c}, SPDC downconversion probabilities of TOP and BOT sources, with respect to their time delay $\Delta T$. }
    \label{fig:S8}
\end{figure*}

The rate of SPDC events happening from a pulsed laser is proportional to the average power of the pump signal. We can calculate said SPDC probability $P_{SPDC}$ by recording the ratio of the number of coincidences between signals and idlers of the same pump pulse and different pump pulses. The number of coincidences of signal and idler from the same pump pulse, so at $T_D = 0$ is proportional to the time of recording $\tau$, the loss of signals and idlers $L_s$ and $L_i$, and the downconversion probability $P_{SPDC}$
\begin{align}
    C_{T_D=0} &\propto \tau L_i L_s P_{SPDC} \\
    C_{T_D=rn} &\propto \tau L_i L_s P_{SPDC}^2 \\
    D &= \frac{C_{T_D=rn}}{C_{T_D=0}} = P_{SPDC}
\end{align}

for the case of coincidences between different pulses at $T_D = rn$, with $r$ being the repetition rate of the pulsed laser, the number of coincidences is proportional to the time and losses, but in this case to the downconversion probability squared. Thus, by calculating the ratio $D$ between the pulses, we can extract $P_{SPDC}$.

In the performed two source interference experiment, we sweep the position of a fiber collimation package, thereby introducing a sweep in time difference between the pulses. Moving one of the collimation packages also changes the coupling slightly, therefor we record the ratios between the coincidences for each data point in the experiment and then normalize to the change of SPDC probability, to get a normalized four fold coincidence rate. Since the signals of both the sources are coupled via a directional coupler, with reflectivity of about $R=0.625$, we have to take into account coincidences between signals and idlers from different sources but same pulses when doing the calculation of the SPDC probability.
\begin{align}
    C_{T_D=0, i} &\propto \tau L_i L_s P_{SPDC, i} (R + (1-R)P_{SPDC, j}) \\
    C_{T_D=rn, i} &\propto \tau L_i L_s P_{SPDC, i} (R P_{SPDC, i} + (1-R)P_{SPDC, j}) \\
    D &= \frac{C_{T_D=rn, i}}{C_{T_D=0, i}} = \frac{R P_{SPDC, i} + (1-R)P_{SPDC, j}}{R + (1-R)P_{SPDC, j}}
\end{align}
where $\{i,j\}\in \{\mathrm{TOP, BOT}\}$. The calculated SPDC probabilities from the TOP and BOT sources at different stage positions can be seen in Fig. \ref{fig:S8}(c), where Fig. \ref{fig:S8}(a) and (b) show the bidirectional histogram of the outputs of each of the sources, at the stage set to $\Delta T = 0$. We see that the TOP source probability does vary quite significantly, which is explained due to different coupling efficiency depending on the stage position. The BOT source probability varies, as expected, significantly less over the long measurement duration.

\end{document}